\documentclass[prb,showpacs,preprintnumbers,amsmath,amssymb,twocolumn]{revtex4-1}
\usepackage{graphicx}
\usepackage{dcolumn}
\usepackage{bm}
\usepackage{epstopdf}

\begin{document}

\title{Complex magnetism and strong electronic correlations in Ce$_2$PdGe$_3$}

\author{R. E. Baumbach$^{1,4}$, A. Gallagher$^1$, T. Besara$^1$, J. Sun$^{2,3}$, T. Siegrist$^{1,2}$, D. J. Singh$^3$, J. D. Thompson$^4$, F. Ronning$^4$ and E. D. Bauer$^4$}
\address{$^1$National High Magnetic Field Laboratory, Florida State University, Tallahassee, FL}
\address{$^2$FAMU-FSU College of Eng., Dept. Chem. \& Biomed. Eng., Tallahassee, FL}
\address{$^3$Oak Ridge National Laboratory, Oak Ridge, TN}
\address{$^4$Los Alamos National Laboratory, Los Alamos, NM}
\date{\today}

\begin{abstract}
Single crystal x-ray diffraction, magnetic susceptibility, heat capacity, and electrical resistivity measurements are reported for the new tetragonal compound Ce$_{2}$PdGe$_{3}$, which forms in the space group $P4_{2}/mmc$ (\#131) $-$ a relative of the $\alpha$-ThSi$_2$-type structure. Measurements reveal a two-part antiferromagnetic phase transition at $T_{\rm{N,1}}$ $=$ 10.7 K and $T_{\rm{N,2}}$ $=$ 9.6 K and subsequent ferromagnetic ordering near $T_{\rm{C}}$ $\approx$ 2.25 K. The ordered ground state emerges from a lattice of Ce ions that are hybridized with the conduction electrons, as revealed by the enhanced electronic coefficient of the specific heat $\gamma$ $\approx$ 50 mJ/mol-Ce-K$^2$ (extrapolated to $T$ $=$ 0 for $T$ $<$ $T_{\rm{C}}$). Electronic structure calculations suggest that there is significant f-electron weight in the density of states near the Fermi energy and that the Fermi surface includes sheets with distinct nesting vectors. Chemical/structural disorder also plays an important role, as evidenced by results from single crystal x-ray diffraction, the width of the peaks in the heat capacity at $T_{\rm{N}}$ and $T_{\rm{C}}$, and the small residual resistivity ratio RRR $=$ $\rho_{\rm{300K}}$/$\rho_{\rm{0}}$ $=$ 1.8.
\end{abstract}
Classification: Physical Science: Physics

\maketitle

\section{Introduction}
Cerium based Kondo lattice compounds have received considerable attention, owing largely to the variety of exotic ground states that they exhibit (e.g., unconventional superconductivity, heavy fermion behavior, quantum criticality, etc.).~\cite{Stewart_01,Lohneysen_07,Pfleiderer_09} These phenomena result from coupling between the f- and conduction electron states and mainly derive from the competition between the Kondo effect and the Ruderman-Kittel-Kasuya-Yosida (RKKY) interaction, which vary in relative strength depending on the magnitude of the magnetic exchange interaction $J$.~\cite{Doniach_77,kondo,ruderman,kasuya,yosida} In the simplest picture, the Ce nearest-neighbor distances are the dominant variables which governs $J$. From this point of view, it is useful to study classes of materials where higher order structures are built from simple atomic units and, therefore, distances between atoms are a controllable tuning parameter. In this way, the relationship between crystal chemistry and resulting physical phenomena can be explored.

A well known example is that of BaAl$_4$ (space group $I$4/$mmm$) which is the fundamental unit for a variety of prototypical correlated electron systems, including those that crystallize in the ThCr$_2$Si$_2$, CaBe$_2$Ge$_2$, and BaNiSn$_3$ structures.~\cite{parthe,pearson} This family of materials includes some of the most intensely studied correlated electron systems (e.g., CeCu$_2$Si$_2$,~\cite{assmus_84,yuan_03} CeRh$_2$Si$_2$,~\cite{movshovich_96} YbRh$_2$Si$_2$,~\cite{trovarelli_00,friedemann_09} and URu$_2$Si$_2$.~\cite{palstra_85,schlabitz_86,maple_86,mydosh_14}) Structural trends are also seen for materials based on Cu$_3$Au-type clusters (space group $Pm$3$m$): i.e., the heavy fermion antiferromagnet CeIn$_3$ is the fundamental building block for the layered tetragonal materials Ce$_n$$T_m$In$_{3n+2m}$ ($n$ $=$ 1, 2, ...; $m$ $=$ $n$ = 1, 2, ... and $T$ $=$ transition metal).~\cite{thompson_12} This series includes the extensively studied heavy fermion antiferromagnetic quantum critical point (QCP) systems Ce$T$In$_5$ ($T$ $=$ Co, Rh, and Ir).~\cite{petrovic_01,hegger_00,petrovic_01_b} It is particularly noteworthy that while the cubic CeIn$_3$ and layered versions (e.g., CeRhIn$_5$) exhibit qualitatively similar phase diagrams, the maximum superconducting transition temperature is larger by nearly a factor of ten for the layered versions, suggesting that two dimensional confinement of the correlated quasiparticles is beneficial for amplifying the effects of the QCP.~\cite{mathur_98,hegger_00} Together with expectations from theory,~\cite{monthoux_99} this information provides an important boundary condition for designing correlated electron superconductors with optimized properties.

Herein, we report results for the new Ce-based compound, Ce$_{2}$PdGe$_3$, which forms in a tetragonal structure with space group $P4_{2}/mmc$ (\#131). This structure is derived from that of $\alpha$-ThSi$_2$ and, as a result, bears structural similarities to the noncentrosymmetric materials that form in the LaPtSi-type structure. Electronic structure calculations suggest that there is significant f-electron weight near the Fermi energy and, consequently, suggest heavy fermion behavior. Calculations also show that the Fermi surface is composed of sheets with well defined nesting vectors. Magnetization, heat capacity, and electrical resistivity measurements reveal a two-part transition into antiferromagnetic order at $T_{\rm{N,1}}$ $=$ 10.7 K and $T_{\rm{N,2}}$ $=$ 9.6 K and subsequent ferromagnetic ordering near $T_{\rm{C}}$ $\approx$ 2.25 K. The enhanced value of the electronic coefficient of the specific heat ($\gamma$ $\approx$ 50 mJ/mol-Ce-K$^2$ inside the ordered ground state) confirms that there is appreciable hybridization between the f- and conduction electron states. Finally, x-ray diffraction and electrical resistivity measurements reveal disorder, owing to site interchange between Pd and Ge. We discuss this material in the context of other $\alpha$-ThSi$_2$ related compounds and propose scenarios for inducing enhanced correlated electron behavior.

\section{Methods}
Single crystals of Ce$_{2}$PdGe$_3$ were grown from elements with purities $>99.9$\% in a molten In flux. The reaction ampoules were made by loading the elements into a 2 cm diameter alumina crucible in
the ratio 1(Ce):2(Pd):1(Ge):20(In). The crucible was sealed under vacuum in a quartz tube, heated to 1050 $^{\circ}$C at a rate of 75$^{\circ}$C/hr, kept at this temperature for 24 hours, and then
cooled to 750 $^{\circ}$C at a rate of 4 $^{\circ}$C/hr. After removing the  flux by spinning the ampoules in a centrifuge, single crystal platelets with typical dimensions of several millimeters on a side and 0.5 - 1 millimeter thickness were collected. Similar specimens of the nonmagnetic La$_2$PdGe$_3$ were also produced in the same way.

A single crystal was mounted on a glass fiber for single crystal x-ray diffraction measurement using an Oxford-Diffraction Xcalibur2 CCD system with graphite monochromated Mo K$\alpha$ radiation. A complete sphere of data was collected using $\omega$ scans with 1$^o$ frame widths. The data collection, indexation, and absorption correction were performed using the Agilent CrysAlisPro software,~\cite{CrysAlisPro} and structure refinements and solution were carried out with CRYSTALS.~\cite{Crystals} The position 4i was refined as two atoms, Ge2 and Pd2, with equivalent atomic positions and atomic displacement parameters, but with joint occupancies adding to 1. For subsequent measurements, a single crystal was selected and aligned using a four-axis Enraf Nonius CAD-4 Single Crystal X-Ray Diffractometer. The obtained orientation matrix allowed for an unambiguous determination of the unit cell axes to within a fraction of a degree. A crystallographic information file (CIF) has been deposited with ICSD (CSD\# 427680).

First-principles calculations were performed within the framework
of density functional theory (DFT),~\cite{ISI:A19641557C00018,ISI:A19657000000015}
and used the full-potential linearized augmented plane-wave plus local
orbitals (FP-LAPW+lo) method,~\cite{ISI:A1234567} as implemented
in the WIEN2k code.~\cite{ISI:000177824600013} The Perdew-Burke-Ernzerhof
(PBE) \cite{ISI:A1996VP22500044} form of the generalized gradient approximation
(GGA) was adopted as the exchange-correlation functional.

Magnetization $M(T,H)$ measurements were carried out for temperatures $T$ $=$ 1.8 - 300 K under an applied magnetic field of $H$ $=$ 1 kOe and for 0 $<$ $H$ $<$ 70 kOe at several different $T$ for $H$ applied both parallel ($\parallel$) and perpendicular ($\perp$) to the c-axis using a Quantum Design VSM Magnetic Property Measurement System. The specific heat $C(T,H)$ was measured for $T$ $=$ 0.37 - 20 K and the electrical resistivity $\rho(T,H)$ was measured for $T$ $=$ 0.4 - 300 K using a Quantum Design Physical Property Measurement System.

\section{Results}
\begin{figure}[!tht]
    \begin{center}
        \includegraphics[width=3.5in]{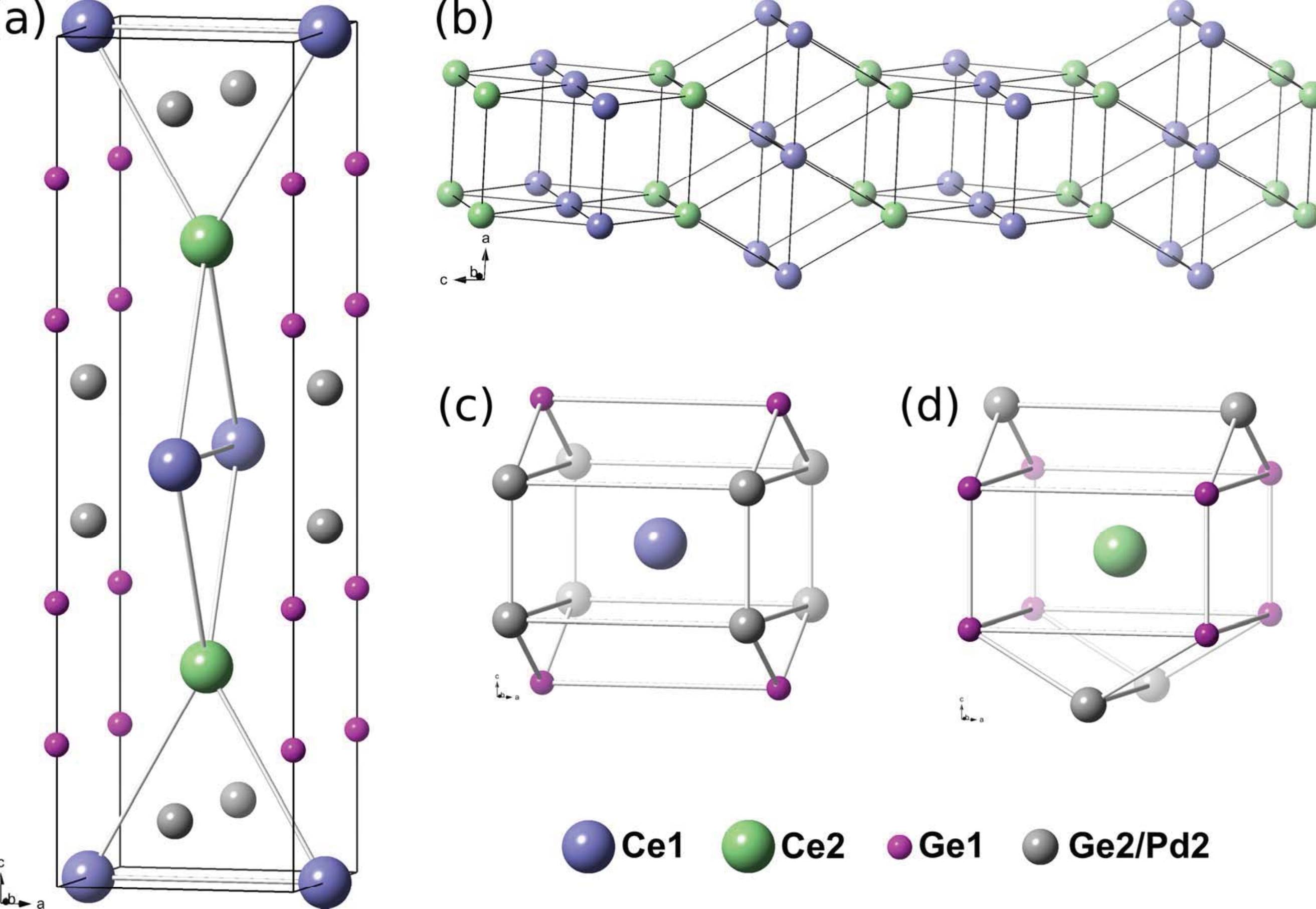}
        \caption{Crystal structure of Ce$_{2}$PdGe$_3$, depicting (a) the unit cell, (b) the array of Ce atoms, (c) local coordination of Ce1, and (d) local coordination of Ce2.}
        \label{fig:struct}
    \end{center}
\end{figure}

Ce$_{2}$PdGe$_3$ crystallizes in the tetragonal space group $P4_{2}/mmc$ (\#131) with unit cell parameters $a=4.24440(8)$~ \textrm{\AA} and $c=14.7928(2)$~\textrm{\AA} (Figure \ref{fig:struct}a). This structure is related to the $\alpha$-ThSi$_2$ type structure with similar structural moieties, but different overall arrangement. This is reflected in the size of the unit cell, which is close in values to the $\alpha$-ThSi$_2$ type structure. Details of the structure measurement and refinement are summarized in Table \ref{tbl:xray}. The structure is comprised of two distinct cerium sites (2c and 2f), one germanium site (4g), and one mixed-occupied germanium/palladium site (4i) with occupancies $\textrm{Ge2}\approx0.458(8)$ and $\textrm{Pd2}\approx0.542(8)$, giving a formal stoichiometry of Ce$_{2}$Pd$_{1.08}$Ge$_{2.92}$. For simplicity, we refer to this compound as Ce$_2$PdGe$_3$ throughout the rest of the manuscript. The atomic positions are summarized in Table \ref{tbl:coord}.

The Ce atoms are arranged in square patterns in the \emph{ab}-plane, and each square array layer consists of either Ce1 or Ce2, with the layers alternating as \ldots Ce1\textendash Ce2\textendash Ce1\textendash Ce2 \ldots along the \emph{c}-axis. The layers, however, are not stacked directly on top of each other, but with the Ce1-layers displaced every other layer along the \emph{a}-axis and every other layer along the \emph{b}-axis with respect to the Ce2-layers. This forms a three-dimensional network of Ce atom arrays in which the Ge1 and mixed Ge2/Pd2 atoms are located. These arrays alternate along the \emph{c}-axis between \ldots AAA\ldots stacking in the \emph{a}-axis and \ldots AAA\ldots stacking in the \emph{b}-axis (Figure \ref{fig:struct}b), and are slightly distorted since the intralayer Ce\textendash Ce distances are 4.244~\textrm{\AA} (identical for both layers), while the interlayer Ce1\textendash Ce2 distances are 4.264~\textrm{\AA}. See Table \ref{tbl:bonds} for a list of bonds.

\begin{table}[!bht]
    \begin{center}
        \caption[]{Selected single crystal x-ray diffraction data, along with collection and refinement parameters.}
        \begin{tabular}{l l}
            \hline
            Compound & Ce$_{2}$Pd$_{1.08}$Ge$_{2.92}$ \\
            \hline
            Formula weight & 607.22 g/mol \\
            Space group & $P4_{2}/mmc$ (\#131)\\
            Unit cell parameters & $a=4.24440(8)$ \textrm{\AA} \\
             & $c=14.7928(2)$ \textrm{\AA} \\
            Volume & 266.491(5) \textrm{\AA}$^{3}$ \\
            $Z$ & 2 \\
            $\rho_{\textrm{calc}}$ & 7.567 g/cm$^{3}$ \\
            Data collection range & $4.8^{\circ}\leq\theta\leq66.53^{\circ}$ \\
            Reflections collected & 1391 \\
            Parameters refined & 13 \\
            $R_{1}$, $wR_{2}$ & 0.0515, 0.0548 \\
            Goodness-of-fit on $F^{2}$ & 0.9982 \\
            \hline
        \end{tabular}
        \label{tbl:xray}
    \end{center}
\end{table}

Both Ce1 and Ce2 are twelve-fold coordinated. Ce1 is surrounded by eight mixed-occupied Ge2/Pd2 atoms forming a square prism, capped by four Ge1 atoms forming two triangle prisms aligned along the \emph{a}-axis (Figure \ref{fig:struct}c). Ce2, on the other hand, has eight Ge1 atoms forming a square prism, while capped by four mixed-occupied Ge2/Pd2 atoms forming two triangle prisms, one along the \emph{a}-axis and the other along the \emph{b}-axis (Figure \ref{fig:struct}d). The Ce1\textendash Ge1 bond is 3.254~\textrm{\AA} while the Ce2\textendash Ge1 bond is slightly shorter, 3.244~\textrm{\AA}. The Ce1\textendash Ge2/Pd2 bond is 3.233~\textrm{\AA} and the Ce2\textendash Ge2/Pd2 3.277~\textrm{\AA}.

\begin{table}[!bht]
    \begin{center}
        \caption[]{Atomic coordinates and equivalent thermal displacement parameters for Ce$_{2}$Pd$_{1.08}$Ge$_{2.92}$ in space group $P4_{2}/mmc$.}
        \begin{tabular}{c c c c c c c}
            \hline
            Atom & Site & Occ. & $x$ & $y$ & $z$ & $U_{\textrm{eq}}$ \\
            \hline
            Ce1 & 2c & 1 & 1/2 & 0 & 1/2 & 0.0066(1) \\
            Ce2 & 2f & 1 & 1/2 & 1/2 & 1/4 & 0.0061(1) \\
            Ge1 & 4g & 1 & 0 & 0 & 0.33323(4) & 0.0093(2) \\
            Ge2 & 4i & 0.458(8) & 0 & 1/2 & 0.41877(1) & 0.0100(1) \\
            Pd2 & 4i & 0.542(8) & 0 & 1/2 & 0.41877(1) & 0.0100(1) \\
            \hline
        \end{tabular}
        \label{tbl:coord}
    \end{center}
\end{table}

As mentioned above, Ce$_2$PdGe$_3$ crystallizes in the space group $P4_{2}/mmc$. This is due to the disordered-mixed-occupied site Ge2/Pd2 and can not be handled directly in an electronic structure calculation. Thus, in order to calculate the density of states, we reconstruct
the structure by labeling all the mixed sites as Ge and subsequently substituting half of them with Pd. Symmetry operations allow only two non-centric subgroup structures in this model, those with the
space groups $P4_{2}mc$ and $P\bar{4}m2$. The x-ray diffraction data refine under both space groups as a twinned crystal with racemic mixture and a Flack parameter $f$ $=0.5$.

For our calculations, the LAPW sphere radii used for both cases were 2.5 Bohr for Ce, 2.0
Bohr for Ge and 2.31 Bohr for Pd. The cut-off parameter for the plane
wave basis was R$_{min}$K$_{max}$=7 and an 8$\times$8$\times$8
k point mesh which has 75 k points in the irreducible Brillouin zone
(IBZ) was used for both structures. We used the experimental lattice
parameters and atomic coordinates which are shown in tables~\ref{tbl:xray} and ~\ref{tbl:coord} and then the atomic coordinates were relaxed with a forced minimization, which is shown in table~\ref{tbl:calc}. The spin-orbit
coupling effect within the second variational procedure is included for all the calculations.

\begin{table}[!bht]
    \begin{center}
        \caption[]{Interatomic distances for Ce$_{2}$Pd$_{1.08}$Ge$_{2.92}$. All distance errors are within 0.001 \AA.}
        \begin{tabular}{l r l r}
            \hline
            Bond & d (\textrm{\AA}) & Bond & d (\textrm{\AA}) \\
            \hline
            Ce1\textendash Ce1     & 4.244 & Ge1\textendash Ge1 (\emph{a}-/\emph{b}-axis) & 4.244 \\
            Ce2\textendash Ce2     & 4.244 & Ge1\textendash Ge1 (\emph{c}-axis) & 2.463 \\
            Ce1\textendash Ce2     & 4.264 & Ge2/Pd2\textendash Ge2/Pd2 (\emph{a}-/\emph{b}-axis) & 4.244 \\
            Ce1\textendash Ge1     & 3.254 & Ge2/Pd2\textendash Ge2/Pd2 (\emph{c}-axis) & 2.402 \\
            Ce1\textendash Ge2/Pd2 & 3.233 & Ge1\textendash Ge2/Pd2 & 2.471 \\
            Ce2\textendash Ge1     & 3.244 &  &  \\
            Ce2\textendash Ge2/Pd2 & 3.277 &  &  \\
            \hline
        \end{tabular}
        \label{tbl:bonds}
    \end{center}
\end{table}

\begin{table}[!bht]
    \begin{center}
        \caption[]{Atomic coordinates of the two ordered crystal structures used in
the calculations.}
        \begin{tabular}{ccccccc}
   \hline
Space group & Atom & Site & Occ. & x & y & z\tabularnewline
   \hline
{$P4_{2}mc$} & Ce1 & 2c & 1 & 1/2 & 0 & 0.50055(4)\tabularnewline
 & Ce2 & 2b & 1 & 1/2 & 1/2 & 0.24938(7)\tabularnewline
 & Ge1 & 2a & 1 & 0 & 0 & 0.33444(1)\tabularnewline
 & Ge2 & 2a & 1 & 0 & 0 & 0.66797(8)\tabularnewline
 & Ge3 & 2c & 1 & 0 & 1/2 & 0.58120(4)\tabularnewline
 & Pd & 2c & 1 & 0 & 1/2 & 0.41645(5)\tabularnewline
   \hline
{$P\bar{4}m2$} & Ce1 & 2g & 1 & 0 & 1/2 & 0.24882(9)\tabularnewline
 & Ce2 & 1a & 1 & 0 & 0 & 0\tabularnewline
 & Ce3 & 1d & 1 & 0 & 0 & 1/2\tabularnewline
 & Ge1 & 2f & 1 & 1/2 & 1/2 & 0.08413(0)\tabularnewline
 & Ge2 & 2f & 1 & 1/2 & 1/2 & 0.58364(5)\tabularnewline
 & Ge3 & 2g & 1 & 1/2 & 0 & 0.33022(5)\tabularnewline
 & Pd & 2g & 1 & 1/2 & 0 & 0.16598(6)\tabularnewline
   \hline
\end{tabular}
        \label{tbl:calc}
    \end{center}
\end{table}

\begin{figure}[!tht]
    \begin{center}
        \includegraphics[width=3.2in]{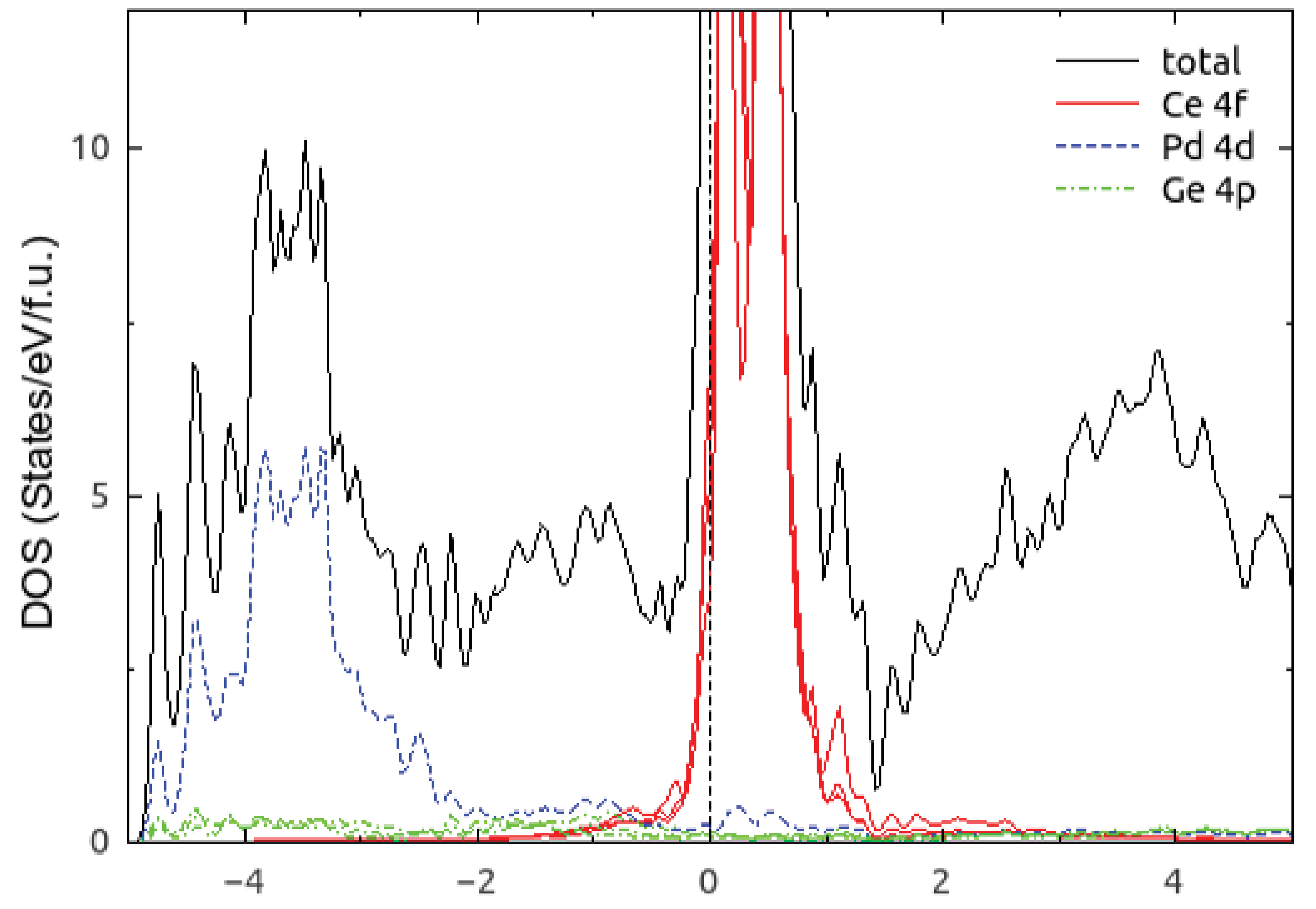}
        \caption{Calculated total and projected density of states of Ce$_{2}$PdGe$_{3}$}.
        \label{fig:DOS}
    \end{center}
\end{figure}

Starting from the two ordered structures, we find that the one
with space group $P\bar{4}m2$ has fewer free atomic position parameters than
the $P4_{2}mc$ one, as shown in table~\ref{tbl:calc}. Our
nonmagnetic DFT calculations also show a relatively lower total energy
of about 0.03 eV/atom, similar total density of states (TDOS), and projected density of
states (PDOS) as the first case. Therefore, we take the $P$$\bar{4}$$m2$ case as
the representative to discuss the TDOS and PDOS results, which are
shown in Fig.~\ref{fig:DOS}.

\begin{figure}[!tht]
    \begin{center}
        \includegraphics[width=2.8in]{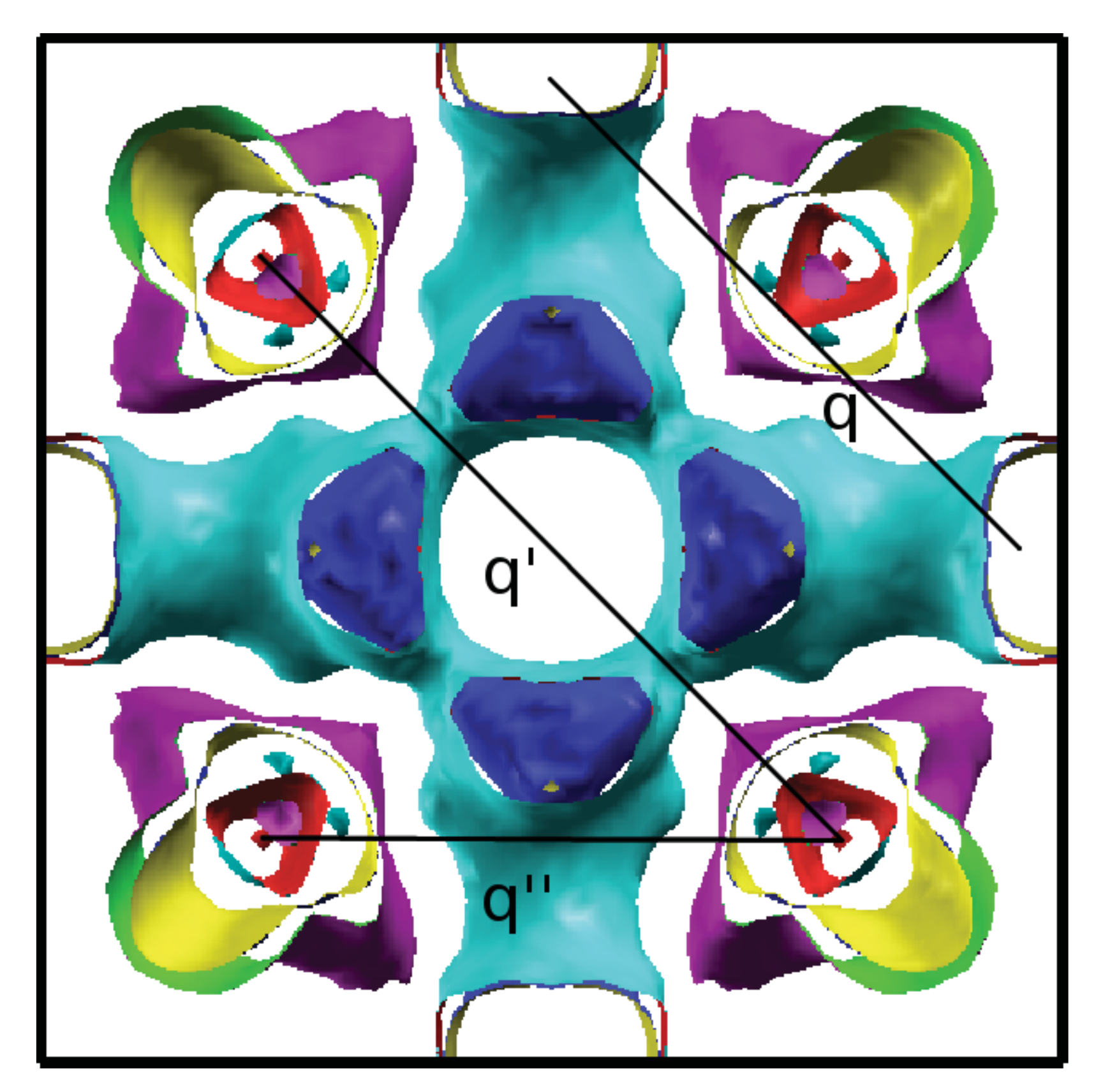}
        \caption{Calculated Fermi surface of Ce$_{2}$PdGe$_{3}$. Three different nesting vectors are shown by q, q', and q''.}
        \label{FS}
    \end{center}
\end{figure}

Note that there are three Ce-atomic sites in the $P\bar{4}m2$ case, as shown in table~\ref{tbl:calc}. This results from the splitting of one Ce site (Ce 2c) in the original Pd/Ge mixed structure with space group $P$4$_2$/$mmc$ when the symmetry is lowered to $P$$\bar{4}$$m2$. But the PDOS results show that the density of states of these three Ce sites closely overlap with each other, and can be considered as two different Ce sites, as seen in the experiments. The main contribution to the DOS at
the Fermi level ($E_{\rm{F}}$) stems from the Ce 4f states, which have a value
of 15 states eV$^{-1}$ per formula unit. This result suggests intermetallic heavy fermion behavior with hybridization between the Ce 4f and Pd 4d states and Ge 4p states, as can be seen from the significant overlap in the energy region from E$_{F}$
down to -2 eV. The moderate 4f-4d/4p hybridization indicates that the correlation effects may be weak enough that a typical
GGA method is sufficient to give a reasonable electronic structure.~
\cite{ISI:000182822300155} The f-electron density of states shows a characteristic large peak. Importantly, for this compound the lower part is occupied leading to dominant f-electron character at the Fermi level and an occupation consistent with the experimental observation of Ce f-moments. The Pd d-states are mostly below the Fermi energy and are concentrated in the energy range from -5 eV to -2 eV. The conduction bands above 1.5 eV mainly originate
from Ce 5d states (not shown), as expected. The calculated Fermi surface is shown in Fig.~\ref{FS}, where we find several large sheets. Importantly, there are at least three potentially strong nesting features seen as shown by q, q', and q''.

\begin{figure}[tbp]
    \begin{center}
    \includegraphics[width=3.25in]{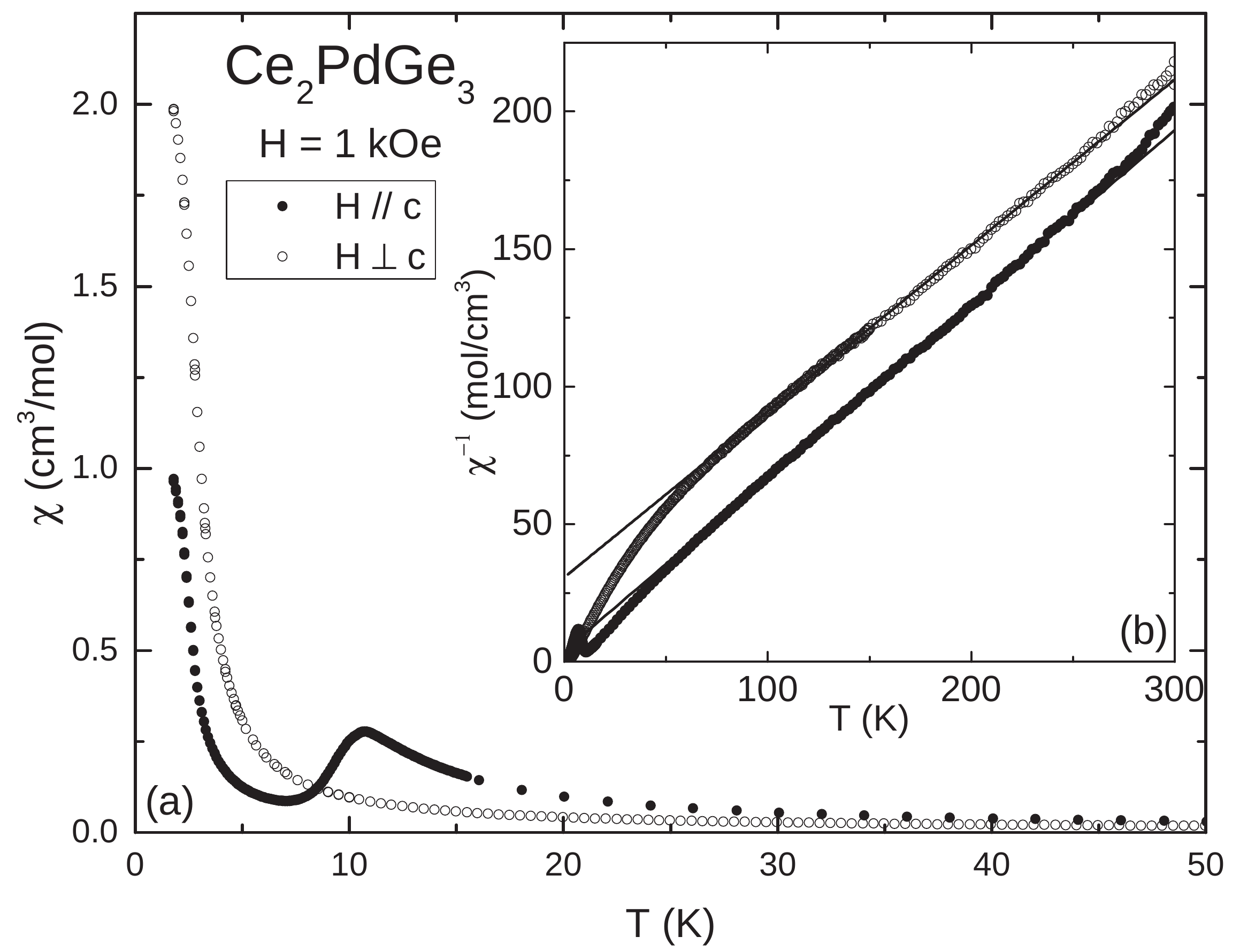}
    \end{center}
    \caption{(a) Magnetic susceptibility $\chi(T)$ $=$ $M(T)/H$  for $H$ $=$ 1 kOe applied parallel ($\parallel$) and perpendicular ($\perp$) to the c axis vs. temperature $T$ for Ce$_2$PdGe$_3$. (b)
 $\chi^{-1}(T)$ for $H$ $\parallel$ and $\perp$ c. The solid lines are Curie-Weiss fits to the data, as described in the text.}
    \label{M_T}
\end{figure}

The magnetic susceptibility $\chi(T)$ $=$ $M(T)/H$ data for $H$ $=$ 1 kOe applied parallel ($\parallel$) and perpendicular ($\perp$) to the c axis are summarized in Fig.~\ref{M_T}.  As shown in
Fig.~\ref{M_T}b, Curie-Weiss (CW) behavior given by the expression,
\begin{equation}
\chi = C/(T-\Theta)
\label{CW}
\end{equation}
is observed for 100 K $\leq$ $T$ $\leq$ 300 K, where $\Theta$ $=$ -6.3 K for $H$ $\parallel$ c (-50.9 K for $H$ $\perp$ c), indicating antiferromagnetic correlations with magnetocrystalline anisotropy. The Curie constants
$C$ yield effective magnetic moments $\mu_{eff}$ $\approx$ 2.51 $\mu_B$/Ce (2.57 $\mu_B$/Ce) for $H$ $\parallel$ ($\perp$) c, which are close to what is expected for localized Ce$^{3+}$ ions ($\mu_{\rm{eff}}$ $=$ 2.54 $\mu_{\rm{B}}$/Ce). As temperature
decreases, the behavior of $\chi(T)$ is consistent with splitting of the Hund's rule multiplet by the crystalline electric field (CEF). At low temperatures, complicated magnetic ordering is observed which reflects the presence of two distinct Ce sites in the structure. For $H$ $\parallel$ c, $\chi(T)$ evolves through a peak near $T_{\rm{N,1}}$ $=$ 10.7 K, indicating the onset of antiferromagnetic ordering along the c-axis, and a broad shoulder additionally appears near $T_{\rm{N,2}}$ $=$ 9.6 K. The susceptibility subsequently begins to increase below $T$ $\approx$ 6.9 K and goes through an inflection point near $T_{\rm{C}}$ $=$ 2.25 K which suggests that some fraction of the Ce ions order ferromagnetically below this temperature. For $H$ $\perp$ c, we find no evidence for antiferromagnetism, but $\chi(T)$ again increases with decreasing $T$ and undergoes an inflection point near $T_{\rm{C}}$ $=$ 2.25 K.

\begin{figure}[tbp]
    \begin{center}
    \includegraphics[width=3.25in]{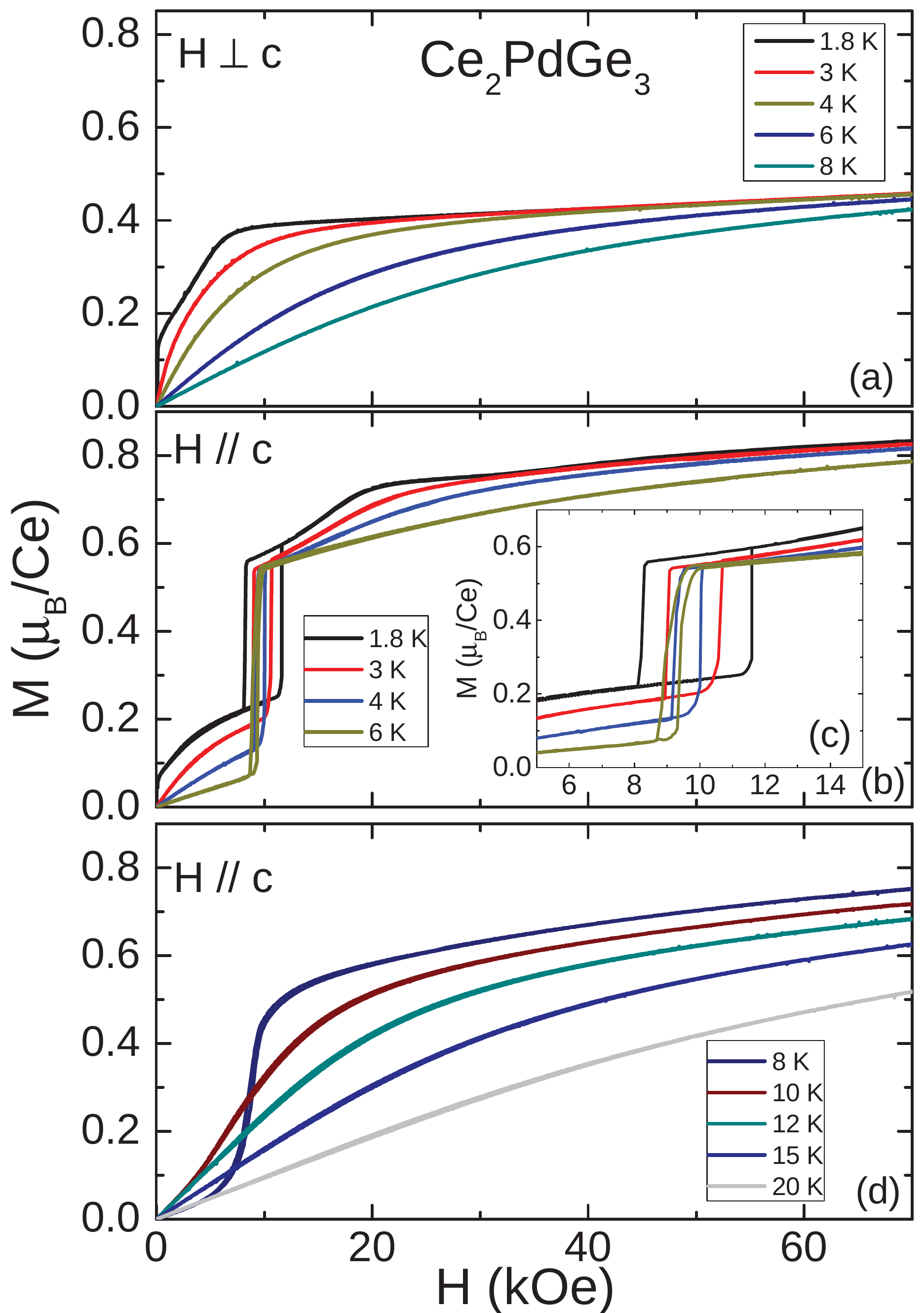}
    \end{center}
    \caption{(a) Magnetization $M$ vs. magnetic field $H$ applied perpendicular to the c-axis for Ce$_2$PdGe$_3$. (b) $M(H)$ for $H$ applied parallel to the c-axis in the low temperature range 1.8 K $<$ $T$ $<$ 6 K. (c) Close-up of the hysteresis region $H_{\rm{M}}$ = 8-11 kOe. (d) $M(H)$ for $H$ applied parallel to the c-axis in the intermediate temperature range 8 K $<$ $T$ $<$ 20 K.}
    \label{M_H}
\end{figure}

In order to further elucidate the magnetic behavior, we show plots of $M(H)$ at various temperatures for $H$ $\parallel$ ($\perp$) c in Fig.~\ref{M_H}. As shown in Fig.~\ref{M_H}a, $M(H)$ for $H$ $\perp$ c exhibits ferromagnetic behavior for $T$ $<$ 3 K where the spontaneous ordered moment $M_{\rm{so}}$ $=$ 0.14 $\mu_B$/Ce and the saturation moment is near $M_{\rm{sat}}$ $\approx$ 0.4 $\mu_B$/Ce above $H$ $\approx$ 6.7 kOe. The value of $M_{\rm{sat}}$ is roughly half of what is expected for a $\emph{s}$ $=$ 1/2 Ce ion in a doublet ground state. This may indicate that the ferromagnetic ordering for this direction is confined to one of the two Ce sites. For $T$ $\geq$ 3 K, we no longer observe a spontaneous ordered moment, consistent with the observation that the ferromagnetic ordering onsets around $T_{\rm{C}}$ $=$ 2.25 K. Moreover, we find that the saturation moment remains roughly 0.4 $\mu_B$/Ce, revealing that it is difficult to polarize about half of the Ce ions in the ab plane, even for $T$ $>$ $T_{\rm{C}}$.

We additionally find complicated behavior for $H$ $\parallel$ c. As shown in Fig.~\ref{M_H}b, $M(H)$ exhibits ferromagnetic behavior for $T$ $<$ 3 K where the spontaneous ordered moment at $T$ $=$ 1.8 K is $M_{\rm{so}}$ $=$ 0.06 $\mu_B$/Ce. Again, no spontaneous ordered moment is observed for $T$ $\geq$ 3 K. We find that for $T$ $<$ 10 K, $M(H)$ abruptly increases (with hysteresis) between $H_{\rm{M}}$ $\approx$ 8 - 11 kOe (Fig.~\ref{M_H}c). With increasing $T$, the value of $H_{\rm{M}}$ slightly decreases while the width of the hysteresis loop decreases dramatically and disappears for $T$ $>$ 8 K. At low temperatures, we also find a well defined shoulder around 20 kOe which broadens and disappears with increasing $T$. Similar behavior is seen in a variety of other heavy fermion systems, both with and without magnetic order (e.g., in CeSi$_{1.81}$~\cite{drotziger}) and has been associated with a weakening of the Kondo effect. Above 20 kOe, $M_{\rm{sat}}$ approaches the the expected value for a $s$ $=$ 1/2 Ce ion in a doublet ground state, revealing that all of the Ce ions can be polarized in this direction. Finally, for $T$ $>$ 10 K, $M(H)$ recovers Brillouin-like behavior (Fig.~\ref{M_H}d), consistent with the observation that the system undergoes its initial magnetic ordering near $T_{\rm{N}}$ $\approx$ 10.7 K.

\begin{figure}[tbp]
    \begin{center}
    \includegraphics[width=3.25in]{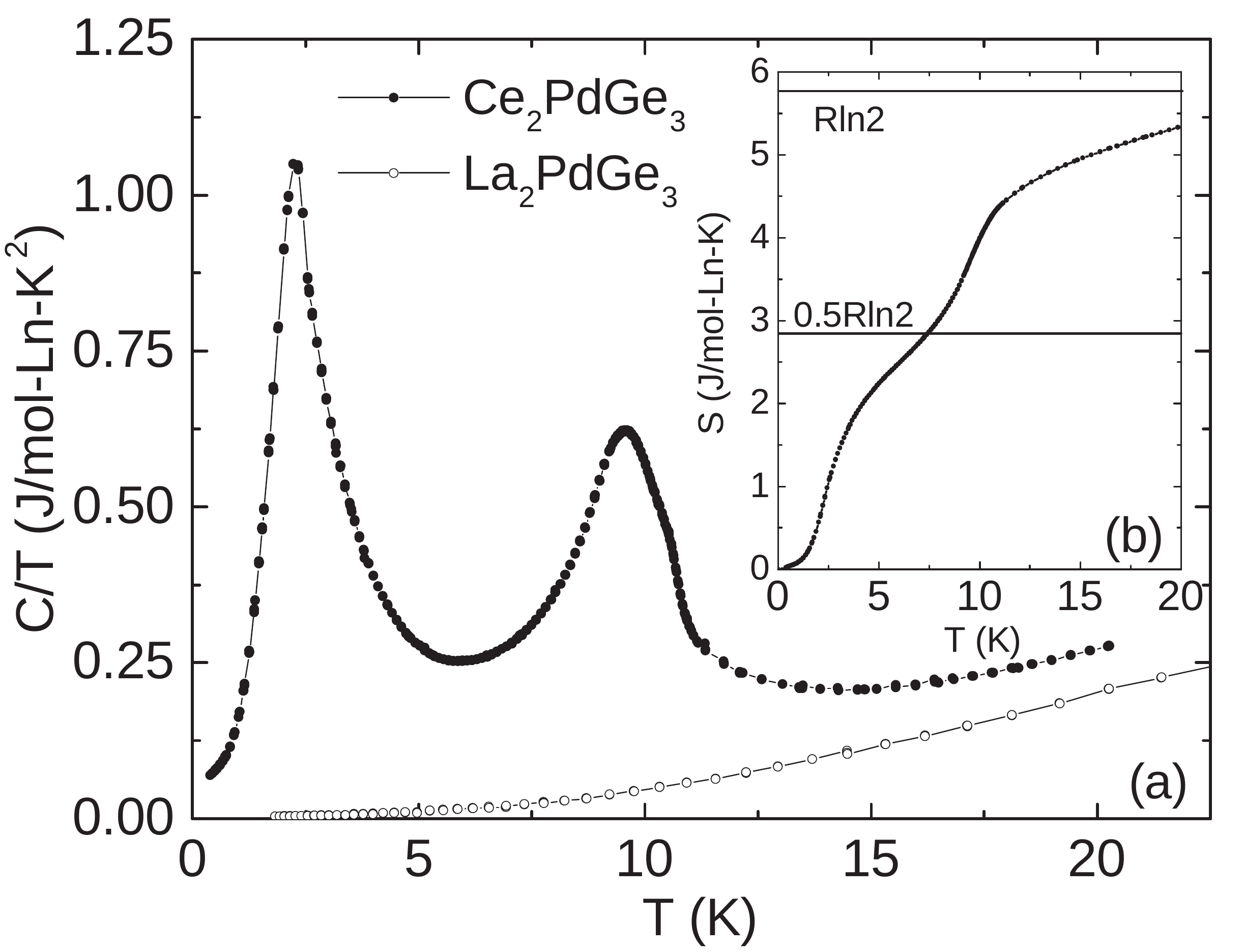}
    \end{center}
    \caption{(a) Heat capacity divided by temperature $C/T$ vs $T$ for Ce$_2$PdGe$_3$ and La$_2$PdGe$_3$. (b) 4f contribution to the entropy $S$ vs. $T$ for $H$ $=$ 0 for Ce$_2$PdGe$_3$, calculated as described in the text.}
    \label{C}
\end{figure}

Heat capacity measurements for La$_2$PdGe$_3$ and Ce$_2$PdGe$_3$ are shown in Fig.~\ref{C}. La$_2$PdGe$_3$ exhibits behavior that is typical for a nonmagnetic metal, where the data are described for 1.8 K $<$ $T$ $<$ 9 K by the expression $C/T$ $=$ $\gamma$ $+$ $\beta$$T^2$. Here, we find that the electronic coefficient $\gamma$ $=$ 3.8 mJ/molK$^2$ and the phonon coefficient $\beta$ $=$ 0.74 mJ/molK$^4$, corresponding to a Debye temperature $\theta_{\rm{D}}$ = 250 K. In contrast, Ce$_2$PdGe$_3$ shows behavior that indicates magnetism in the presence of strong electronic correlations. We find two convoluted humps, at $T_{\rm{N,1}}$ $=$ 10.7 K and $T_{\rm{N,2}}$ $=$ 9.6 K, consistent with results from $\chi(T)$. A third feature appears as a broadened peak at $T_{\rm{C}}$ $=$ 2.25 K, reflecting the onset of ferromagnetism seen in $\chi(T)$. We further note that the peak shapes are somewhat broad, possibly due to disorder arising from the Pd-Ge site interchange. Finally, the value of $\gamma$, extrapolated from within the ordered state, is roughly 50 mJ/mol-Ce-K$^2$. We emphasize that both in the paramagnetic and ordered states, $C/T$ is enhanced by comparison to the nonmagnetic La-analogue.

In Fig.~\ref{C}b we show the 4f contribution to the entropy, which was obtained by integrating $C/T$ after subtracting the La$_2$PdGe$_3$ data and extrapolating to zero temperature. Here, we find that 0.77Rln2/Ce is recovered near 10.7 K and $S(T)$ extrapolates towards Rln2/Ce above 20 K. This result is consistent with the Ce ions being in a doublet ground state that is produced by crystal electric field splitting of the Hund's rule multiplet, where the Kondo driven hybridization between the f- and conduction electron states is strong enough to remove a significant fraction of the 4f entropy. Similar behavior is seen in a variety of Ce-based correlated electron systems that exhibit Kondo compensated ordered moments: e.g., as for such prototypical systems as CeRhIn$_5$ and CePd$_2$Si$_2$.~\cite{hegger_00,dhar}  We additionally note that the magnetic entropy is nearly evenly distributed between the two magnetic transitions.

\begin{figure}[tbp]
    \begin{center}
    \includegraphics[width=3.25in]{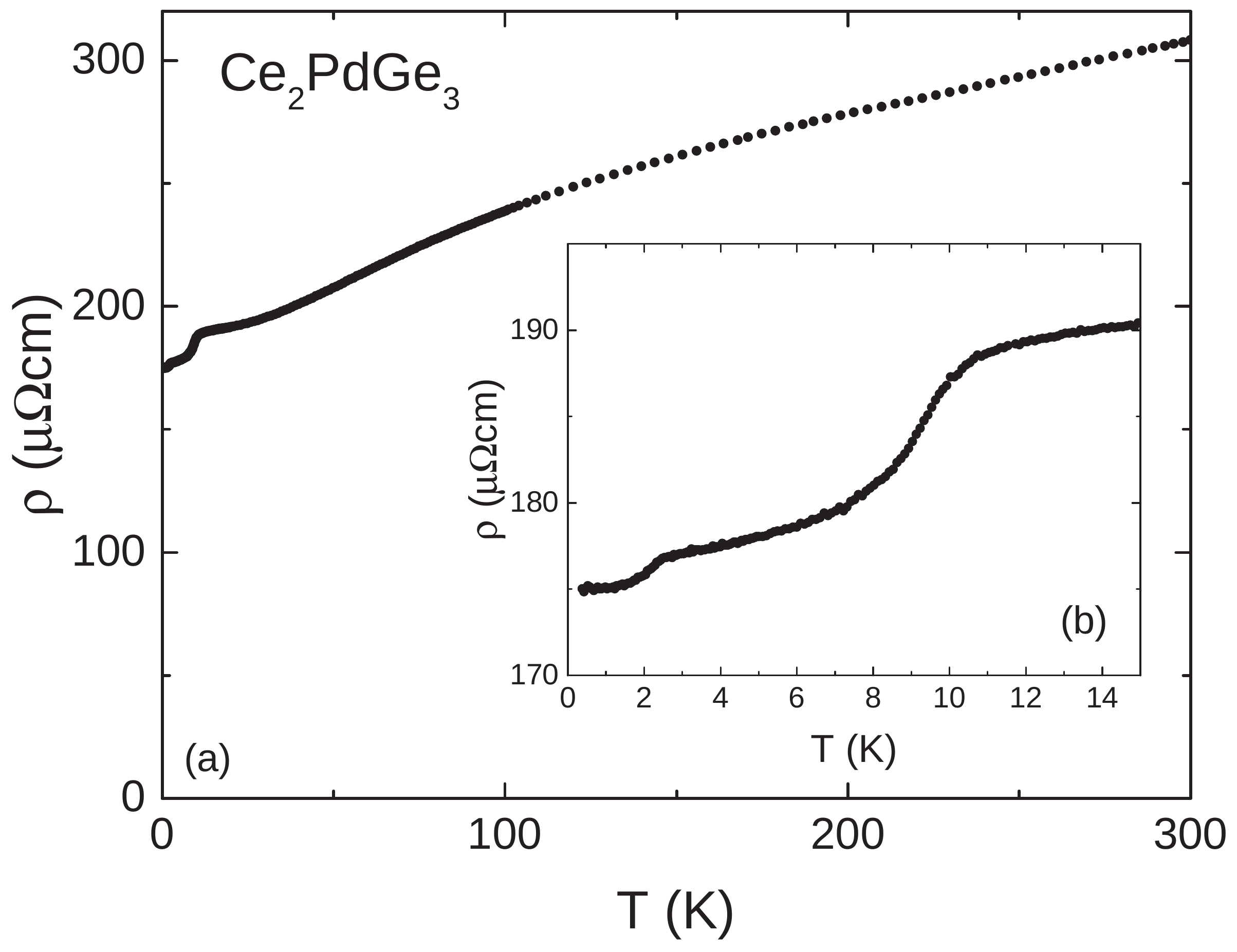}
    \end{center}
    \caption{(a) Electrical resistivity $\rho$ vs. temperature $T$ for Ce$_2$PdGe$_3$. (b) Low temperature zoom of $\rho(T)$. }
    \label{R}
\end{figure}

The electrical resistivity $\rho(T)$ data, where the current was applied in the ab-plane, are shown in Fig.~\ref{R}. For temperatures above 100 K, $\rho(T)$ weakly decreases and exhibits a broad shoulder. In Ce-based compounds, this type of behavior is typically due to a combination of the onset of Kondo coherence on the lattice of Ce ions and
a depopulation of the higher crystal electric field split levels. As the temperature decreases further, $\rho(T)$ goes through two sharp decreases at the ordering temperatures (Fig.~\ref{R}b), in
agreement with $\chi(T)$ and $C(T)$. We interpret this behavior as being due to a reduction of the spin disorder scattering at $T_{\rm{N}}$ and $T_{\rm{C}}$.  $\rho(T)$ finally saturates near 175 $\mu$$\Omega$cm, giving a residual resistivity
ratio RRR $=$ $\rho_{300K}$/$\rho_0$ $\approx$ 1.8. The value of RRR is consistent with a significant amount of disorder scattering due to the site interchange between the Pd and Ge ions.

\section{Discussion}
Taken together, our results show that Ce$_2$PdGe$_3$ exhibits strong electronic correlations that arise from hybridization between the Ce f-electron and conduction electron states. Electronic band structure calculations reveal significant f-electron weight near the Fermi energy and the electronic coefficient of the specific heat ($\gamma$ $\approx$ 50 mJ/mol-Ce-K$^2$) supports this point of view. We additionally find complicated magnetic ordering at low temperatures: in small fields, a two part evolution into an antiferromagnetically ordered state occurs at $T_{\rm{N,1}}$ $=$ 10.7 K and $T_{\rm{N,2}}$ $=$ 9.6 K. This is followed by ferromagnetic ordering below $T_{\rm{C}}$ $\approx$ 2.25 K. We note that ferromagnetic ordering is uncommon in Ce-based compounds, making Ce$_2$PdGe$_3$ a member of a small group which includes CeRu$_2$Ge$_2$,~\cite{wilhelm_1999_1} CeRh$_3$B$_2$,~\cite{cornelius_94}, CeRu$_2$$M_2$X ($M$ $=$ Al, Ga and $X$ $=$ B, C),~\cite{baumbach_12,matsuoka_12,baumbach_12_1,baumbach_12_2} and CeRuPO.~\cite{krellner_07} In principle, this makes Ce$_2$PdGe$_3$ an attractive candidate to search for a ferromagnetic quantum phase transition. However, the large degree of disorder on the mixed Pd-Ge site presents a significant complication.

Measurements of magnetic isotherms reveal additional interesting behavior, which is likely related to there being two Ce sites. For $H$ $\leq$ 10 kOe, the magnetic ions are most easily polarized for $H$ $\perp$ c-axis. However, near 10 kOe $M(H)$ for $H$ $\parallel$ c undergoes a hysteretic (first order) transition into a more strongly polarized state and subsequently goes through another broad increase with a shoulder near $H$ $=$ 20 kOe (possibly due to a weakening of the Kondo effect), finally saturating towards a value near 0.8 $\mu_B$/Ce. In contrast, when $H$ $\perp$ c, $M(H)$ saturates towards a value of 0.4 $\mu_B$/Ce which suggests that only one of the two species of Ce can be polarized in this direction. Neutron diffraction measurements will be useful to clarify the magnetic structure in these various regimes.

\begin{figure}[tbp]
    \begin{center}
    \includegraphics[width=3.25in]{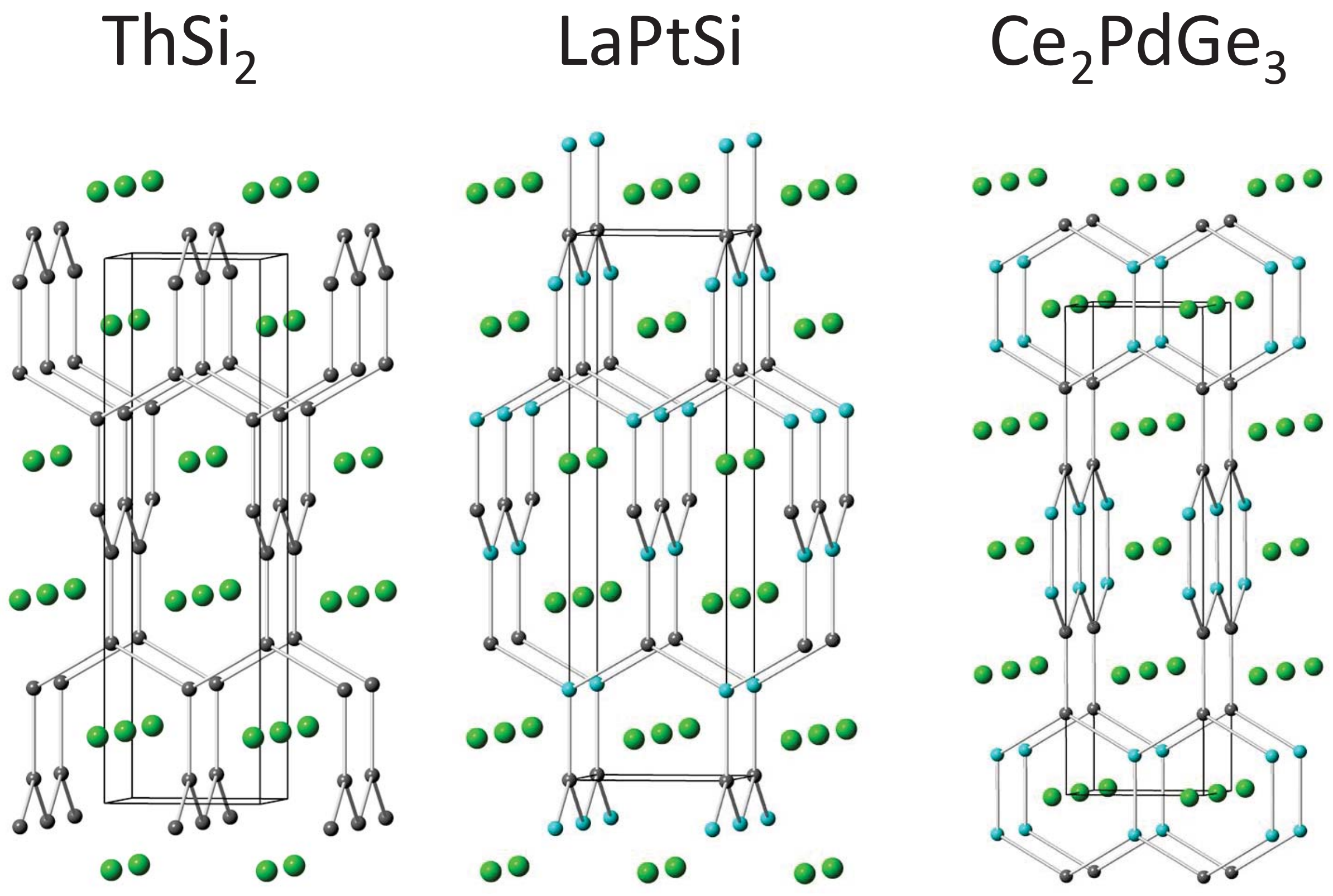}
    \end{center}
    \caption{Left: The tetragonal $\alpha$-ThSi$_2$ structure. Center: The noncentrosymmetric LaPtSi structure. Right: The Ce$_2$PdGe$_3$ structure.}
    \label{ThSi2}
\end{figure}

Our understanding of this compound is partially clarified by our calculations, which reveal several sheets with two dimensional character (Fig.~\ref{FS}). There are three distinct nesting vectors that indicate a tendency towards incommensurate antiferromagnetic order, which may be expected to compete with the observed low temperature ferromagnetism and are the probable origin of the observed antiferromagnetism above $T_{\rm{C}}$. We note that nesting induced antiferromagnetism is in general more sensitive to disorder than itinerant ferromagnetism. As such, the antiferromagnetic tendency may be suppressed by the disorder on the mixed site, and it may be that samples with different levels of disorder would show different transition temperatures between the ferromagnetic and antiferromagnetic states, and perhaps even antiferromagnetism at low $T$ in highly ordered material. In any case, the competition between itinerant antiferromagnetism and ferromagnetism in this heavy electron material implies that interesting signatures of non-trivial quantum fluctuations involving spin degrees of freedom may be expected in clean well ordered samples.

Finally, it is of interest to examine related compounds to search for structural-electronic trends. As mentioned above, Ce$_2$PdGe$_3$ is related to the $\alpha$-ThSi$_2$ structure (Fig.~\ref{ThSi2}:left). We note that the compounds Ce(Si,Ge)$_{2-x}$ and CeSi$_{2-x}$Ge$_x$ crystallize in this form, and exhibit behaviors that can be understood in the context of the Doniach picture, spanning from intermediate valence to heavy fermion ferromagnetism.~\cite{lahiouel,drotziger,yashima} Ternary compounds can also be built from $\alpha$-ThSi$_2$, including those with the LaPtSi-type structure (Fig.~\ref{ThSi2}:middle), which alternates the Pd and Si atoms along the c-axis, with atoms at the same z-coordinate all being the same type. This metal ordering on the Si-lattice of $\alpha$-ThSi$_2$ results in the loss of the center of inversion, making LaPtSi a non-centric intermetallic compound with space group $I4_{1}md$. While the close chemical relatives CePdSi and CePdGe do not form in this structure,~\cite{lipatov,seropegin} there are two related examples CeNiSi and CePtSi, which exhibit intermediate valence and nonmagnetic heavy fermion behavior, respectively.~\cite{lee} We also show the structure for Ce$_2$PdGe$_3$ (Fig.~\ref{ThSi2}:right), where the close similarity to $\alpha$-ThSi$_2$ and LaPtSi is evident: in each case, the atomic cluster shown in Fig.~\ref{fig:struct}d is the basic building block. Therefore, given the tendency in this class of materials towards strong Kondo-driven hybridization and ferromagnetism, it may not be surprising that Ce$_2$PdGe$_3$ exhibits the reported behavior. Therefore, it is attractive to infer that related compounds may provide an environment in which to study correlated electron ferromagnetism and, possibly, ferromagnetic quantum criticality.

\section{Conclusion}
We have uncovered the compound Ce$_2$PdGe$_3$ which crystallizes in the space group $P4_{2}/mmc$ (\#131), a derivative of the $\alpha$-ThSi$_2$ structure. Our calculations and measurements show that this compound exhibits Kondo-driven hybridization between the f- and conduction electron states. We find complex magnetic ordering at low temperatures with a two-part antiferromagnetic transition at $T_{\rm{N,1}}$ $=$ 10.7 K and $T_{\rm{N,2}}$ $=$ 9.6 K, and ferromagnetism below $T_{\rm{C}}$ $\approx$ 2.25 K. Magnetic isotherms further reveal complicated behavior, which likely derives from the presence of two Ce sites in the crystal structure. Neutron diffraction measurements are needed to clarify the magnetic structure. Finally, it might be of interest to search for a quantum phase transition, either through chemical substitution, applied pressure, or exploration of structurally related analogues, as FM-QPTs are quite unusual in Ce-based correlated electron systems.

\section{Acknowledgements}
Bulk properties measurements (R.B. and A.G.) were performed at the National High Magnetic Field Laboratory, which is supported by National Science Foundation Cooperative Agreement No. DMR-1157490, the State of Florida, and the U.S. Department of Energy. Electronic structure calculations (J. S. and D. S.) were done at Oak Ridge National Laboratory and were supported by the Department of Energy, Basic Energy Sciences, Materials Sciences and Engineering Division and the GO ORNL program. Single crystal XRD measurements (T. B. and T. S.) were supported by the U. S. Department of Energy, Office of Science, Basic Energy Science, under award DE-SC0008832. Synthesis of materials and preliminary magnetization measurements (R.B., J.T., F. R., and E. B.) were performed under the auspices of the US Department of Energy, Office of Basic Energy Sciences, Division of Materials Sciences and Engineering, and PECASE funding from the US DOE, OBES, Division of Material Science and Engineering.

\end{document}